\documentclass[12pt,a4paper]{article}

\usepackage[english]{babel}
\usepackage[T2A]{fontenc}
\usepackage[cp1251]{inputenc}
\usepackage{amsmath}
\usepackage{graphicx}
\usepackage{amssymb}
\usepackage{color}
\usepackage{amsfonts}
\usepackage{wrapfig}
\usepackage{caption}

\usepackage[hyphens,spaces,obeyspaces]{url}
\usepackage[colorlinks,breaklinks,allcolors=blue]{hyperref}
\usepackage{doi}

\usepackage{physics}
\usepackage{siunitx}

\begin{document}
	
	\binoppenalty=10000
	\relpenalty=10000
	
\begin{center}
	\textbf{\Large{Triple point in a cell fluid model with effective temperature-dependent attraction}}
\end{center}

\vspace{0.3cm}

\begin{center}
 M.P.~Kozlovskii, O.A.~Dobush\footnote{e-mail:  dobush@icmp.lviv.ua}, R.V.~Romanik, I.V.~Pylyuk, M.A.~Shpot
\end{center}

\begin{center}
	Yukhnovskii Institute for Condensed Matter Physics \\ National Academy of Sciences of
	Ukraine \\ 1, Svientsitskii Street, 79011 Lviv, Ukraine
\end{center}

	 \vspace{0.2cm}

	\begin{center}
 \small \textbf{Abstract}
	\end{center}
\noindent	\small
We study a cell fluid model of a many-particle system with Curie-Weiss-type interaction potential. It is considered as an open system in a fixed volume partitioned into a large number of congruent cubic cells. The interaction potential comprises two competing components: a global uniform attraction acting between all particle pairs in the volume and a short-range repulsion between particles occupying the same cell. Previous studies have established that this model admits an exact solution, exhibits multiple critical points, and undergoes a sequence of first-order phase transitions. Despite variations in the interaction strengths, no triple point appears as long as these parameters remain fixed. We demonstrate that incorporating effective {temperature-dependent} attractive interactions fundamentally alters the phase behavior of the cell model. This modification preserves the model's exact solvability while resulting in the emergence of a triple point in the phase diagram.


	\normalsize
\section{Introduction}

Studies of special lines and special points in thermodynamic phase diagrams of many-particle systems are of central interest for statistical physics (see e.g., \cite{Stanley71,CL95,PathriaBeale21}). Physical systems like the ubiquitous water can typically exist in a solid, liquid, and gas-like (vapour) states. The latter two states, distinct from solid, are often termed as \emph{fluid}.

Typical phase diagrams of such systems feature the lines of coexistence where different pairs of phases can be simultaneously present in the system. In the case of water, the coexistence line of the liquid and vapour phases has the important and well-known reference point --- the boiling point at the one-atmosphere pressure ($P$) and temperature $t=\SI{100}{\celsius}$. On the high-temperature side, this coexistence line ends with the \emph{critical point} (at $t=t_c\simeq\SI{374}{\celsius}$ and $P=P_c\simeq\SI{218}{atm}$) where the coexisting liquid and vapor phases become identical and the transition order changes from first-order (discontinuous) to second-order (continuous).

Close to the usual freezing temperature of water at $t=\SI 0{\celsius}$, and namely at $t=t_{tr}\simeq\SI{0.01}{\celsius}$ and $P=P_{tr}\simeq\SI{0.006}{atm}$, \emph{three} phases (solid, liquid and vapour) with three different densities $\rho_S$, $\rho_L$, and $\rho_V$ are able to coexist. This special point on the phase diagram is called the \emph{triple point}. A detailed pressure-temperature phase diagram of water can be found in \cite{WikiTP}.

In the pressure-temperature phase diagram, the triple point corresponds to the unique intersection of the three coexistence lines: the solid-liquid (fusion) line, the liquid-vapour (boiling) line, and the solid-vapour (sublimation) line, all of them being the first-order phase transition lines. Due to the differences in densities of coexisting phases, the triple point cannot be represented as a single point in the temperature-density phase diagram --- see \cite[p. 2]{HansenMcDonald13}. There, the density $\rho_{tr}$ corresponds to the liquid-phase density $\rho_L$, while the other two densities $\rho_S$, and $\rho_V$ define the endpoints of the three-phase coexistence region at the triple-point temperature $T=T_{tr}$. This picture is in full qualitative agreement with our theoretical finding presented in Sec.~\ref{sec:tp}.

Finalizing this physical part of Introduction, we stress the fundamental distinction between a triple and \emph{tricritical point}. At the triple point three first-order transition lines meet and the three involved phases coexist retaining their individual densities. By contrast, the tricritical point is an end-point of a three-phase coexistence region, where all three coexisting phases become simultaneously identical and the phase transition changes its order from first to second \cite{LawrieSarbach84}. By the Gibbs phase rule, triple points can exist in one-component fluids like water, while tricritical points are only accessible in multi-component systems \cite{Widom96} such as that used in the classical experiment described in \cite[p. 6-7]{LawrieSarbach84}.

The extent of theoretical investigation of special points mentioned above, varies considerably. Critical points have been the subject of intensive study for over 150 years, for a nice historical review see \cite{BHK09}; some modern references may include \cite{Fis74,DGL,Goldenfeld,BDFN,PV02}.

Tricritical points have received substantial but more limited attention since the middle of the last century. An interesting history of the development of understanding the tricritical point can be found in \cite[Sec.~II]{KnoblerScott84}.
The theoretical development can be traced by following the exposition in \cite{LawrieSarbach84} and references quoting this classical review article.

The situation with the triple point seems to be rather specific. In the past, this special point in the phase diagram played a crucial role in experimental studies of the absolute temperature scale (see \cite{WikiTP}). Also, as mentioned in \cite{WikiTP}, the triple point has been known since 1871 under this same name.
Surprisingly, the literature on the theory of triple points appears to be quite limited.

Early attempts to describe a triple point in the framework of lattice gas theories have been made in \cite{WB77,Bell80}. However, classical lattice gas models \cite{LY52} are not able to incorporate a triple point because there appears only one first-order phase transition in such systems. By contrast, double-occupancy~\cite{LYZ21} or multiple-occupancy~\cite{FHL04,FL18} lattice gases may be expected to exhibit triple points as they possess more complicated phase diagrams.

In this paper, we consider a cell fluid model introduced in~\cite{KKD18,KKD20}, which represents an example of a multiple-occupancy system, and suggest its modification that leads to an emergence of a triple point. In this modification, each stable phase is characterized by a distinct temperature-dependent attraction strength. By assigning different temperature dependencies of the attraction coupling constant to each phase, we enable a qualitatively new thermodynamic behavior while maintaining the mathematical structure of the original formulation.

In \cite{KKD18,KKD20}, the inter-particle interaction has been originally assumed to be temperature independent. With this assumption, the cell fluid model exhibits multiple critical points and a sequence of first-order phase transitions~\cite{KD22,DKPP26}. However, because all phases are governed by the same interaction strength, this model cannot support the coexistence of more than two phases simultaneously and therefore is unable to exhibit a triple point. This limitation stands in contrast to real fluids and soft-matter systems, where triple points exist and are extensively studied in experiments and computer simulations~\cite{WikiTP,Young2023book,AhmedSadus09,DeitersSadus22,Azhar00,Tuinier08}.

A natural motivation for extending the original model definition arises from the observation that coarse-grained or effective particle interactions often explicitly depend on temperature~\cite{Sobolewski2012,Abbott15,HohmZarkova04}. Many materials display a strengthening of effective attractive interactions upon cooling, particularly in dense environments or near structural transitions. These considerations suggest that incorporating temperature-dependent attraction may be a minimal physically motivated way to enrich the phase behavior of the cell fluid model while preserving its analytical solvability.

The paper is organized as follows. In Section~\ref{sec:model}, we briefly recall the definition and main results for the cell fluid model. In Section~\ref{sec:temp-dep1}, we introduce its modification and define the effective temperature-dependent attraction strengths assigned to each phase. In Section~\ref{sec:noname} we point out the new features in the solution for the grand partition function, obtain explicit formulas for the chemical potential, density and pressure, and in Section~\ref{sec:coexistence} analyze the coexistence conditions for possible pairs of phases. In Section~\ref{sec:tp}, we present the piecewise equation of state and the pressure–temperature phase diagram, highlighting the emergence of a critical point and a triple point. Final discussion on the role of the proposed modification of the cell fluid model is presented in Conclusions.

\section{\label{sec:model}Model definition and phase diagram}

This study is based on a multiple-occupancy cell fluid model with Curie–Weiss-type interactions (CFM), originally introduced in~\cite{KKD18} and further developed in~\cite{KKD20,KD22}. The model describes an open system of interacting point particles in a three-dimensional volume $V$ partitioned into $N_v$ mutually disjoint, congruent cubic cells $\Delta_l$, $l\in\{1,...,N_v\}$, of volume $v$.
The pairwise Curie-Weiss interaction potential $\Psi_{N_v} (x,y)$ between particles with coordinates $x$ and $y$ reads~\cite{KKD20,KD22}:
\begin{equation}\label{0d2}
	\Psi_{N_v} (x, y)=- \frac{J_1}{N_v} + J_2 \sum_{l =1}^{N_v}
	I_{\Delta_l}(x) I_{\Delta_l}(y).
\end{equation}
The first term in~\eqref{0d2} represents a global mean-field–type attraction of strength $J_1>0$ acting between any pair of particles in the system. The second term with the coupling constant $J_2>0$ represents a local repulsion between particles located in the same cell $\Delta_l$, while the indicator function $I_{\Delta_l}(x)$ is defined as
\begin{equation}	\label{0d3}
	I_{\Delta_l}(x)=\left\{
	\begin{array}{l}
		1, x\in \Delta_l , \\
		0, x\notin \Delta_l.
	\end{array}
	\right.
\end{equation}
The interaction potential~\eqref{0d2} satisfies the Ruelle's thermodynamic stability condition~\cite{Ruelle70} provided that $J_2>J_1>0$.
\begin{figure}[h!]
	\centering
	\includegraphics[width=0.6\textwidth]{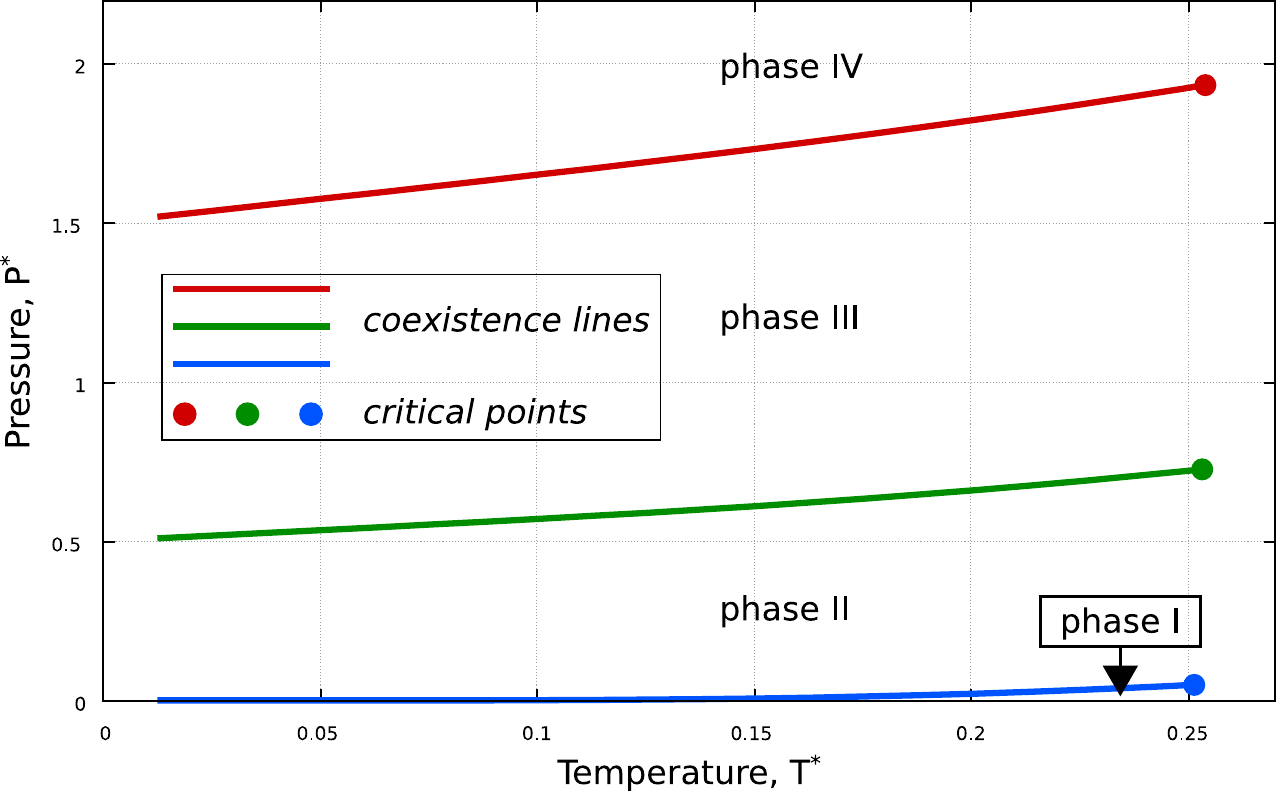}
\caption{Phase diagram in the pressure-temperature plane for the cell fluid model with Curie-Weiss-type interaction potential~\eqref{1d1} and repulsion-to-attraction ratio $f=J_2/J_1=1.5$. The first three phase coexistence lines (blue, green, and red) are shown, which separate four distinct stable phases (I -- IV) and correspond to first-order phase transitions between these phases. Each coexistence line terminates at its critical point, indicated by a solid circle. 	}\label{fig1}
\end{figure}

As reported in~\cite{KKD18,KKD20,KD22}, the model admits an exact solution in the thermodynamic limit, and, at sufficiently low temperatures exhibits an infinite sequence of first-order phase transitions between phases of increasing density.
A typical pressure-temperature phase diagram of the CFM is shown in Figure~\ref{fig1}, where $T^*=k_{\rm B} T/J_1$ is the dimensionless temperature, $T$ being the absolute temperature and $k_{\rm B}$ the Boltzmann constant, and $P^*=P v/J_1$ is the dimensionless pressure. The plots show the phase equilibria in the $(P^*, T^*)$-projection. The temperatures span from as low as $T^* =0.01 $ up to the supercritical region. The figure features first three phase coexistence lines, which correspond to a sequence of first-order phase transitions: the blue line represents the transition between Phase~I and Phase~II, the green line indicates the transition between Phase~II and Phase~III, and the red line corresponds to the transition between Phases~III and~IV. Each coexistence line terminates at its respective critical point, which results in an infinite set of critical points.

The values of some important physical quantities at the first three critical points are presented in Table~\ref{tab1}. In addition to the critical temperature $T^*_c$ and pressure $P^*_c$, the Table~\ref{tab1} contains numerical data for the dimensionless chemical potential $\mu^*=\mu/J_1$ ($\mu$ being the physical chemical potential) and the dimensionless particle number density $\rho^*=\langle N\rangle/N_v$ (where $\langle N\rangle$ is the mean particle number in the system and $N_v$ the number of cells) related to these critical points.
\begin{table}[h!]
		\caption{Dimensionless parameters of the critical points: the critical temperature $T^*_c$, particle number density at criticality $\rho^*_c$, critical pressure $P^*_c$ and chemical potential $\mu^*_c$. The table includes data related to the first three critical points numbered by $n=1,2,3$. Numerical values are obtained for the ratio $f=1.5$ and dimensionless cell volume $v^*=5$ (for explanation of $v^*$ see Section~\ref{sec:noname}).}
	\tabcolsep4.5pt
	\label{tab1}
\begin{center}
		\begin{tabular}{|c|c|c|c|c|}
		\hline
		$n$	&   $T^{*(n)}_c$    &   $\rho^{*(n)}_c$     &  $P^{*(n)}_c$     &   $\mu^{*(n)}_c$    \\
		\hline
		1	&     0.251285  &   0.503869   &    0.0488583  &    0.365215  \\
		\hline
		2	&   0.253109    &    1.50136   &   0.725778  &     1.03938 \\
		\hline
		3	&   0.253838   &    2.50070   &   1.93280  &     1.64220 \\
		\hline
	\end{tabular}
\end{center}
\end{table}

Notably, as it is shown in~\cite{DKPP26}, changes in the magnitude of the repulsion-to-attraction ratio $f := J_2/J_1$ do not result in emergence of triple points in the phase diagrams of the CFM with repulsion and attraction strengths taken as temperature independent constants.

\section{\label{sec:temp-dep1} Incorporating the effective temperature-dependent attraction}
A novel feature of the current paper is that we phenomenologically incorporate an effective temperature-dependent attraction strength into the model. Thus, we assume that:
\begin{enumerate}\itemsep-1mm
\item The inter-particle attraction strength $J_1$ in \eqref{0d2} becomes temperature-dependent at sufficiently low temperatures $T^*<T^{*(1)}_c$.
\item The attraction increases as the temperature drops.
\item The attraction's temperature dependence varies between different phases. Within Phase I, it remains constant, as before. At any fixed temperature  in the range $T^*<T^{*(1)}_c$, each subsequent phase, i.e. Phase II, Phase III, etc.,
    is characterized by a stronger attraction compared to a preceding one.
\end{enumerate}
Thus, following the above prescriptions we express the attraction part of interaction as follows, where we use, for brevity, the short-hand notation $T^{*(1)}_c=T^*_c$:
\begin{equation}\label{1d1}
J_\pi(T^*)=\left\{
\begin{array}{l}
	J_1, \quad T^* \geq T^*_c,
	\\
    J_1 \varphi_{\rm \pi}(T^*), \quad T^*<T^*_c\quad\mbox{and}\quad\pi=II, III, \ldots.
\end{array}
\right.
\end{equation}
The temperature dependence of the coupling constant $J_\pi(T^*)$ is introduced through the function $\varphi_{\pi}(T^*)$, specific for each phase with $\pi\ne$ I, via
\begin{equation}\label{1d2}
\varphi_{\pi}(T^*)=\left(\frac{T^*}{T^*_c}\right)^{-\alpha_{\pi}}\equiv
\Theta^{-\alpha_{\pi}},
\qquad 0<\alpha_{II}<\alpha_{III}<\alpha_{IV}<\ldots\,,
\end{equation}
where
\begin{equation}\label{1d2t}
\Theta=\frac{T^*}{T^*_c}<1
\end{equation}
is the \emph{reduced temperature} to be extensively used in the following.
By construction, for $\pi=I$ we have $\varphi_I(T^*)=1$ and $\alpha_I=0$. Other particular values of $\alpha_{\pi}$ will be chosen later.
\\[0.1mm]

With the above extension, the CFM's interaction potential \eqref{0d2} becomes
\begin{equation}\label{1d3}
	\Phi_{N_v} (x, y)=- \frac{J(T^*)}{N_v} + J_2 \sum_{l =1}^{N_v}
	I_{\Delta_l}(x) I_{\Delta_l}(y).
\end{equation}
This modification of the interaction potential does not affect the calculation of the grand partition function, as it does not involve the particle coordinates. Consequently, general results derived in~\cite{KKD20,KD22} remain applicable to the case of temperature-dependent attraction. The only change concerns the form of the coupling constant $J(T^*)$, defined in~\eqref{1d1}. The repulsive interaction between particles within the same cell remains temperature-independent.

\section{\label{sec:noname} Grand partition function and equation of state}

In this section, we briefly outline the main results for the CFM following \cite{KKD20} and \cite{RDKPS25arxiv}.
Its grand partition function is given by 
\begin{equation}\label{1d4}
	\Xi=\sum_{N=0}^\infty \frac{\zeta^N}{N!} \int \limits_V \mathrm dx_1...
	\int \limits_V \mathrm dx_N \exp\left[ -\frac{\beta}{2} \sum_{x,y\in\gamma_N} \Phi_{N_v} (x,y)\right],
\end{equation}
where $\zeta =\exp(\beta\mu)/\Lambda^3$ is the activity, $\mu$ is the chemical potential, $\beta=(k_BT)^{-1}$ is the inverse temperature, $\Lambda= (2\pi\beta\hbar^2/m)^{1/2}$ is the de~Broglie thermal wavelength, $\hbar$ is the reduced Planck constant, and $m$ is the particle mass.
Also in~\eqref{1d4}, $\gamma_N=\{x_1,...,x_N\}$ is a configuration of $N$ particles in the volume $V$, and $x_i$ denotes the space position of the $i$-th particle.

The two-particle interaction potential $\Phi_{N_v}(x, y)$ is given in \eqref{1d3}, where the temperature-dependent attraction coupling constant $J(T^*)$ is defined in \eqref{1d1} and \eqref{1d2}. As in Section~\ref{sec:model}, the magnitude of the function $J(T^*)$ is still controlled by $J_1$, and we continue to use it in rendering all relevant physical quantities dimensionless.

The integral representation of the grand partition function \eqref{1d4} derived in \cite[Sec.~2]{KKD20} and \cite[Sec.~2]{KD22} is given by \cite[(15)]{KD22} and read:
\begin{equation}
	\Xi = (N_v T^*/(2\pi))^{1/2} \int\limits_{-\infty}^{\infty} \mathrm d z \exp(N_v E(T^*,\mu^*;z)).
	\label{1d5}
\end{equation}
The function $E=E(T^*,\mu^*;z)$ is given by
\begin{equation}
	\label{1d6}
	E(T^*,\mu^*;z)=- \frac{T^*}{ 2  \varphi_{\pi}(T^*)}\left(z-\frac{\mu^*}{T^*}- \ln v^*-\frac{3}{2}\ln T^{*}\right)^2 + \ln K_0(T^*;z),
\end{equation}
where $K_0(T^*; z)$ is the $0$-th member of the family of special functions
\begin{eqnarray}\label{def:Kj}
	K_j(T^*;z)=\sum_{n=0}^{\infty} \frac{n^j}{n!} \exp\left(zn- \frac{f}{2T^*}n^2\right).
\end{eqnarray}
In \eqref{1d6}, the dimensionless cell volume $v^*={v}/{\Lambda^3_J}$ appears, where the temperature-independent constant $\Lambda_J=(2\pi\hbar^2/mJ_1)^{1/2}$ is introduced, related to de Boglie thermal wavelength $\Lambda$ via $\Lambda_J=(T^*)^{1/2}\Lambda$. We also recall that the parameter $f$, which appears in the functions \eqref{def:Kj}, is the repulsion-to-attraction ratio $f=J_2/J_1$.

In the thermodynamic limit $N_v \to \infty$ implying $V \to \infty$ with $V/N_v=v=const$, the grand partition function has the asymptotic form~\cite{KKD20,KD22,RDKPS25arxiv}
\begin{equation}
	\label{eq:Xi1}
	\Xi(T^*,\mu^*)\propto\exp \left[ N_v E(T^*, \mu^*;\bar{z}_{\rm max})\right],
\end{equation}
where we omit a negligible constant factor in this regime.

The argument $\bar{z}_{\rm max}=\bar{z}_{\rm max}(T^*,\mu^*)$ of the function $E$ in \eqref{eq:Xi1} appears as the position of its global maximum satisfying simultaneously the couple of conditions
\begin{eqnarray}
	\label{eq:max_cond}
	\frac{\partial E(T^*,\mu^*;z)}{\partial z}\bigg|_{z =\bar{z}_{\rm max}}=0
	\qquad\mbox{and}\qquad
	\frac{\partial^2 E(T^*,\mu^*;z)}{\partial z^2}\bigg|_{z =\bar{z}_{\rm max}}<0
\end{eqnarray}
along with the requirement
\begin{equation}\label{REQ}
E(\bar{z}_{\rm max})>E(z),\qquad\forall z\,\in(-\infty,\infty)\,,
\end{equation}
related to the application of Laplace's method in evaluation of the integral \eqref{1d5} in the thermodynamic limit.

More generally, we shall use the notation $\bar{z}$ to denote any solution of the first condition in Eq.~\eqref{eq:max_cond}, regardless of whether it corresponds to a local maximum, a local minimum, or a global maximum. Thus, $\bar{z}_{\rm max}$ represents a subset of $\bar{z}$ that satisfies the second condition in Eq.~\eqref{eq:max_cond} and corresponds to the global maxima relevant for thermodynamically stable states. In all other instances, the extremum position $\bar{z}$ may also be related to metastable or unstable states, see Refs.~\cite{KD22} and~\cite{RDKPS25arxiv}.

The relation between $\mu^*$ and $\bar{z}$ resulting from the first, extremum condition in~\eqref{eq:max_cond}, is given by
\begin{equation}\label{1d10}
	\mu^* (T^*;\bar{z})=T^*\bar{z}- \varphi_\pi(T^*) \frac{K_1(T^*;\bar{z})}{K_0(T^*;\bar{z})} -T^* \ln v^*-\frac{3}{2} T^* \ln T^{*}.
\end{equation}
If, in addition, the second condition in~\eqref{eq:max_cond} is satisfied, and the corresponding solution $\bar z$ yields the highest value $E(\bar z)$ of $E(z)$, we have to deal with a global maximum of this function, and $\bar z$ has to be identified as $\bar{z}_{\rm max}$.

In this case, the standard thermodynamic definition for the average number of particles $\langle N \rangle=\beta^{-1}\partial\ln\Xi/\partial\mu$ can be applied, and this yields the particle number density $\rho^*=\langle N \rangle / N_v$ as
\begin{equation}\label{1d11}
\rho^*(T^*;\bar{z}_{\rm max})=\frac{K_1(T^*;\bar{z}_{\rm max})}{K_0(T^*;\bar{z}_{\rm max})}\equiv M_1(T^*;\bar{z}_{\rm max})\,,
\end{equation}
where we identify the ratio $K_1/K_0$ in the middle of \eqref{1d11} with the first moment of the discrete Gauss-Poisson distribution implied in \eqref{def:Kj} (cf. \cite[(25)]{KD22} and \cite{DSh24}).

In other words, the definition $\rho^*=\langle N \rangle/N_v$ means that $\rho^*$ has equally the meaning of the mean cell occupancy. This is a crucial physical interpretation of $\rho^*$ for the CFM.

Provided that the conditions \eqref{eq:max_cond} -- \eqref{REQ} are satisfied, we substitute the chemical potential $\mu^*$ from \eqref{1d10}, taken at $\bar z=\bar{z}_{\rm max}$, along with \eqref{1d11}, into~\eqref{1d6}. This eliminates the explicit dependence on $\mu^*$ from~\eqref{1d6}, which transform into
\begin{equation}\label{EF}
	E(T^*;\bar{z}_{\rm max})=\ln K_0(T^*;\bar{z}_{\rm max})-\frac{\varphi_\pi(T^*)} {2T^*}\left[M_1(T^*;\bar{z}_{\rm max})\right]^2.
\end{equation}
The grand partition function \eqref{eq:Xi1} becomes thus
\begin{equation}\label{GGP}
	\Xi(T^*,\mu^*)\propto\exp\left\{ N_v E\left[ T^*;\bar{z}_{\max}(T^*,\mu^*) \right]\right\}.
\end{equation}
The last two equations go in parallel with those in \cite[(26), (27)]{KD22}, apart from the appearance of the phenomenological temperature-dependent function $\varphi(T^*)$ in \eqref{EF}. Its presence does not essentially influence the treatment of the grand partition function \eqref{GGP}, but opens a new possibility in altering the pase behavior of the CFM.

It is well known that the CFM exhibits a hierarchy of first-order phase transitions at sufficiently low temperatures. The coexistence lines between any pair of adjacent phases terminate at corresponding critical points. The critical temperatures $T^{*(n)}_c$ and the corresponding values of the global maximum positions $\bar{z}_{\rm max}(T^{*(n)}_c)\equiv\bar{z}_{c}^{(n)}$ for the $n$-th critical point are obtained as solutions of the system of equations
\begin{equation}
	\left\{
	\begin{array}{l}
		\cfrac{\partial \mu^*(T^*; z)}{\partial z}=0, \\[3mm]
		\cfrac{\partial^2\mu^*(T^*; z)}{\partial z^2}=0.
	\end{array}
	\right.
	\label{1d12}
\end{equation}

The coexistence points between neighboring phases are determined by solving the system of equations
\begin{equation}\label{1d15}
	\left\{
	\begin{array}{l}
		E(T^*;z_1)=E(T^*;z_2), \\[2mm]
		\mu(T^*;z_1)=\mu(T^*;z_2),
	\end{array}
	\right.
\end{equation}
where $z_1$ and $z_2$ are the arguments for two distinct phases (e.g., Phase I and Phase II) at the same temperature and pressure.
Equations~\eqref{1d15} ensure the equilibrium condition for these two coexisting phases.

Using the grand partition function \eqref{GGP} in the standard thermodynamic relation $P V=k_{\rm B} T \ln \Xi$, the equation of state can be written in dimensionless form as
\begin{equation}\label{1d16}
	P^*=T^* E(T^*;\bar{z}_{\rm max}).
\end{equation}
Combined equations~\eqref{1d11} and~\eqref{1d16} yield a parametric representation of the equation of state in pressure–temperature–density coordinates with $\bar{z}_{\rm max}$ serving as the parameter. The chemical potential~\eqref{1d10} can be expressed analogously as a function of the density and temperature.

In the CFM, the order parameter is the difference in densities $\rho^*$ between coexisting phases. A first-order phase transition is characterized by a discontinuous jump in $\rho^*$ at fixed temperature and pressure (or chemical potential).
The coexistence conditions~\eqref{1d15} guarantee equal pressures (via $\text{\eqref{1d16}}$) and chemical potentials in the two coexisting phases.
Critical points correspond to the termination of the coexistence lines, where the first and second derivatives of $\mu^*$ with respect to $z$ vanish simultaneously (see \eqref{1d12}). The corresponding solutions for $T^*$ and $\bar{z}_{\rm max}$ determine the critical temperature, density, chemical potential, and pressure through relations~\eqref{1d10},~\eqref{1d11}, and~\eqref{1d16}.

\section{\label{sec:coexistence} Phase-dependent interaction and phase coexistence}

The main feature of our phenomenological assumption in Sec.~\ref{sec:temp-dep1} concerning the temperature dependence of the effective attraction in the system is that the interaction becomes stronger as the temperature decreases. This implies that each subsequent phase with a bigger density has a stronger inter-particle attraction as a previous one. A simple parametrization proposed for such a dependence is given in equations \eqref{1d1} -- \eqref{1d2}.

\subsection{Implications of a power-law temperature dependence}\label{SPS}

\begin{figure}[t!]
	\centering
	\includegraphics[width=0.6\textwidth]{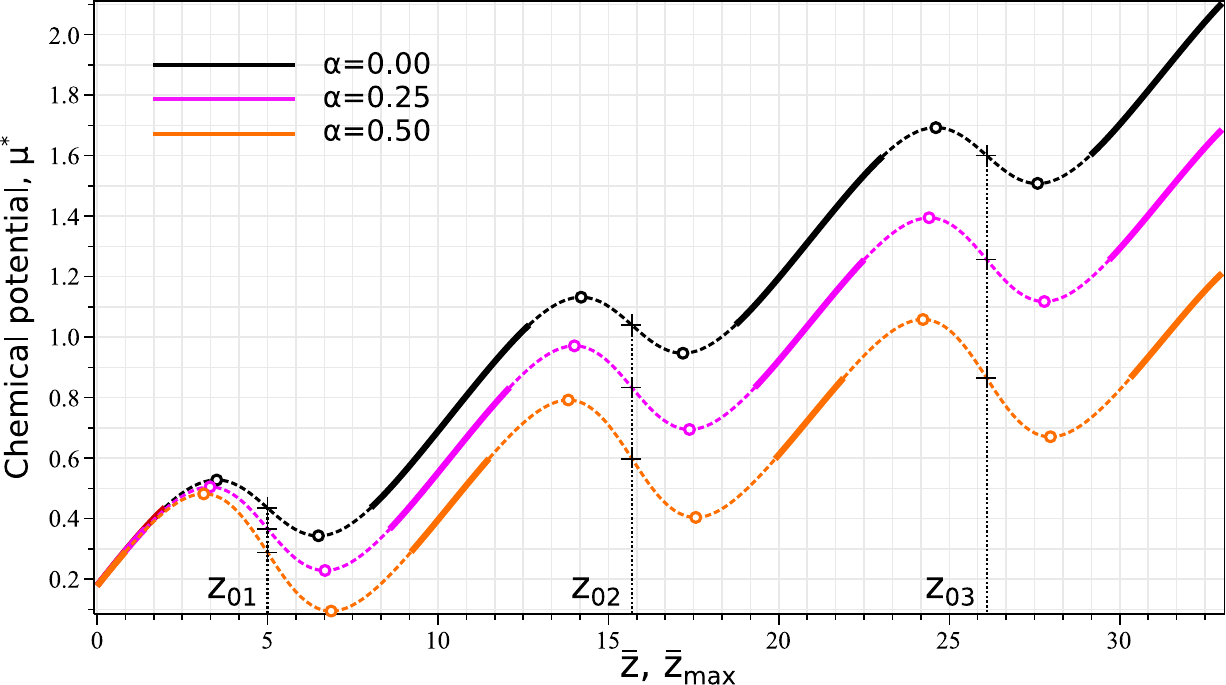}
	\caption{Chemical potential $\mu^*(T^*;\bar z)$ as a function of $\bar z$ given by \eqref{1d10} at fixed value of subcritical temperature $T^*=0.15$ and repulsion to attraction ratio $f=1.5$. The black, magenta, and orange curves correspond to $\alpha=0$, $\alpha=0.25$, and $\alpha=0.5$ in \eqref{1d2}, respectively.
Open circles  denote extrema of $\mu^*$. Black crosses mark geometric midpoints $z_{01}$, $z_{02}$, and $z_{03}$ on segments between nearest maxima and minima pairs.  Bold solid sections of plots show the regions of thermodynamically stable phases given by
$\mu^*(T^*;\bar{z}_{\rm max})$.}
\label{fig2}
\end{figure}

The power-law temperature dependence of the attraction coupling constant $J_\pi(T^*)$ defined via \eqref{1d1} -- \eqref{1d2} results into a reduction of both the chemical potential and pressure, particularly in higher-density phases. To visualize this effect we disregard, for time being, the possibility of $J_\pi(T^*)$ to change from phase to phase. Temporarily, we assume that the same power-law temperature dependence $\varphi_\pi(T^*):=\varphi(T^*)$ applies to all phases, and only change the value of its exponent $\alpha_\pi:=\alpha$ in producing the curves in Figure~\ref{fig2}. It provides thus a view on chemical potential subcritical isotherms plotted using \eqref{1d10}, for $J(T^*)\propto\Theta^{-\alpha}$ with three different exponents $\alpha$. The plotted dependence of $\mu^*$ on its argument is formally extended to values of $\bar{z}$ corresponding to metastable and unstable states (thin dashed segments).

The upper plot of $\mu^*$ corresponds to the case of a constant attraction strength with $\alpha=0$ in \eqref{1d2}, which corresponds to the original formulation of \cite{KKD20}. The intermediate curve represents the case $\alpha= 0.25$, and the lowest one corresponds to $\alpha=0.5$.
Bold solid sections of each plot indicate regions of thermodynamically stable phases realized within a certain range of variable $\bar z_{\rm max}$.
Remarkably, the coordinates of geometric midpoints $z_{01}$, $z_{02}$, and $z_{03}$ between each pair of neighboring extrema coincide, regardless of the strength of temperature dependence of the effective attractive interaction.

For subsequent consideration, it is convenient to introduce the following notation: $z_{\pi}^{in}$ denotes the value of $\bar z_{\rm max}$ where the stability of the Phase~$\pi$ sets-in, and $z_{\pi}^{out}$ is the value of $\bar z_{\rm max}$ where its stability terminates. In particular, all possible states of Phase~I occur when $\bar z$ runs through the whole interval $[0,z_{01})$. However, the variable $\bar z_{\rm max}$, related to the stable state, belongs to the interval $[z_{\pi}^{in},z_{\pi}^{out}]$, which is only a part of the range $[0,z_{01})$.

We write $\bar z_{\rm max}\in [z_{I}^{in},z_{I}^{out}] \subseteq [0,z_{01})$ by acknowledging that the allowed values of $\bar z_{\rm max}$ belong to the interval $[z_{\pi}^{in},z_{\pi}^{out}]$ but do not necessarily take on all values from the interval $[0,z_{01})$. With the same caution, we say that for the Phase~II $\bar z_{\rm max}\in [z_{II}^{in},z_{II}^{out}] \subseteq  [z_{01}, z_{02})$, and for the Phase~III we have $\bar z_{\rm max} \in [z_{III}^{in},z_{III}^{out}] \subseteq  [z_{02}, z_{03})$.

\subsection{Phase-dependent attraction}\label{SQS}

We now go back to our main assumption: the attraction's dependence on temperature varies from phase to phase. Thus, we accept that each stable phase has its own attraction strength $J_{\pi}(T^*)$, defined by equations \eqref{1d1} -- \eqref{1d2}, different for different $\pi$ =I, II, III.
To distinguish the three representative cases, we assign a specific exponent $\alpha_{\pi}$ to each phase via
\begin{equation}\label{1d19}
{\alpha_I}=0,\qquad\alpha_{II}=\frac{1-\Theta}{5}\,,\qquad\alpha_{III}=1-\Theta,
\end{equation}
where the reduced temperature variable $\Theta=T^*/T^*_c<1$ as previously defined in \eqref{1d2t}.

In Phase~I, the attraction strength $J_I$ remains constant: $J_I=J_1$.

In Phases~II and III, the coupling constants $J_\pi$ grow with decreasing temperature according to
\begin{equation}\label{1d21}
J_\pi (T^*)=J_1\Theta^{-\alpha_\pi}.
\end{equation}
As the exponents $\alpha_\pi$ in \eqref{1d21} are related  via $\alpha_{III}=5\alpha_{II}$ by the definition \eqref{1d19}, the grows rate of the effective attraction in the Phase III is essentially faster than in Phase II.

To proceed, we define the chemical potential $\mu^*_\pi$ for each stable phase separately, using the definition \eqref{1d10},
\begin{equation}\label{mual}
	\mu^*_\pi (T^*;\bar z_{\rm max})=T^*\bar z_{\rm max}- \varphi_\pi (T^*) M_1(T^*;\bar{z}_{\rm max}) -T^* \ln v^*-\frac{3}{2} \,T^* \ln T^{*},
\end{equation}
where $\pi=$ I,II,III and $\varphi_I(T^*)=1$.
This is equivalent to introducing the piece-wise effective chemical potential $\mu^* (T^*;\bar z_{\rm max})$ that combines these three cases, via
\begin{equation}\label{1d26}
	 \mu^* (T^*;\bar z_{\rm max})=\left\{
	\begin{array}{l}
		\mu^*_{I} (T^*;\bar z_{\rm max}),  \; \text{if} \; \bar z_{\rm max}\in [z_{I}^{in},z_{I}^{out}] \subseteq [0,z_{01}),  \\
		\mu^*_{II} (T^*;\bar z_{\rm max}),  \; \text{if} \; \bar z_{\rm max}\in [z_{II}^{in},z_{II}^{out}] \subseteq  [z_{01}, z_{02}),\\
		\mu^*_{III} (T^*;\bar z_{\rm max}),  \; \text{if} \; \bar z_{\rm max} \in [z_{III}^{in},z_{III}^{out}] \subseteq  [z_{02}, z_{03}).
	\end{array}
	\right.
\end{equation}

By analogy, for the corresponding functions $E_\pi(T^*;\bar z_{\rm max})$ with $\pi=$ I,II,III we write
\begin{equation}\label{1d29}
	E_\pi(T^*;\bar z_{\rm max})=\ln K_0(T^*;\bar z_{\rm max})-\frac{\varphi_\pi (T^*)}{2T^*}\left[M_1(T^*;\bar{z}_{\rm max})\right]^2
\end{equation}
remembering that $\varphi_I(T^*)=1$.

\subsection{Coexistence lines and their interplay}\label{SCS}

Let us consider whether it is possible to find a sequence of first-order phase transitions in such a modified model. The phase coexistence is possible if the condition of equal chemical potentials and pressures for phases in question is satisfied at the phase transition point.

In our case, for a Phase~I -- Phase~II coexistence it is required that some special values $z_{I}^{out}$ and $z_{II}^{in}$ of $\bar z_{\rm max}$ exist that satisfy the system of equations
\begin{equation}\label{1d30}
	\left\{
	\begin{array}{l}
		E_{I}(T^*; z_I^{out})=E_{II}(T^*; z_{II}^{in}), \\[1mm]
		\mu_I(T^*; z_I^{out})=\mu_{II}(T^*; z_{II}^{in}).
	\end{array}
	\right.
\end{equation}
Similarly, Phase~II and Phase~III coexist provided that there exist such special values $z_{II}^{out}$ and $z_{III}^{in}$ of $\bar z_{\rm max}$ that simultaneous conditions \\
\begin{equation}\label{1d31}
	\left\{
	\begin{array}{l}
		E_{II}(T^*;z_{II}^{out})=E_{III}(T^*;z_{III}^{in}), \\[1mm]
		\mu_{II}(T^*;z_{II}^{out})=\mu_{III}(T^*;z_{III}^{in}).
	\end{array}
	\right.
\end{equation}
are satisfied. Under conditions \eqref{1d30}, two end-points of stable phase regions I and II come together, where the stability of the Phase~I terminates ("out") and the stability of the Phase~II sets-in ("in").
Similarly for conditions in \eqref{1d31}: the end of the stability region II ("out") meets the beginning of the stability region III ("in"). Although the phase label II at $E$ and $\mu$ in \eqref{1d30} and \eqref{1d31} is the same, the respective arguments of these functions are different.

\begin{figure}[h!]
	\centering
	\includegraphics[width=0.32\textwidth]{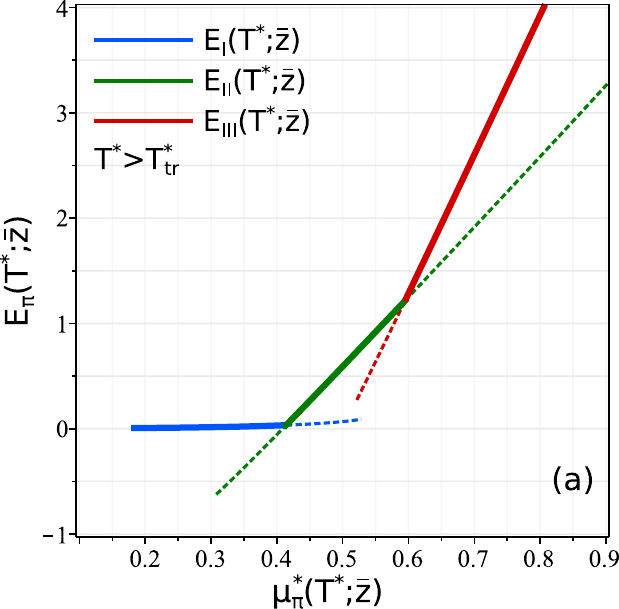}
	\includegraphics[width=0.32\textwidth]{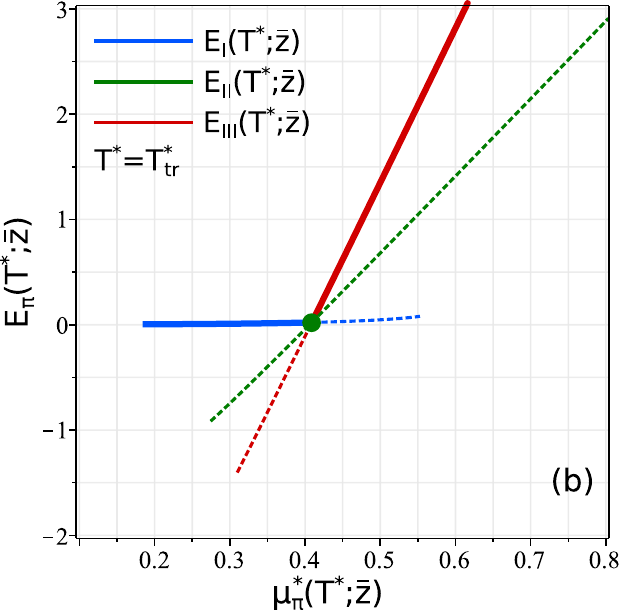}
	\includegraphics[width=0.32\textwidth]{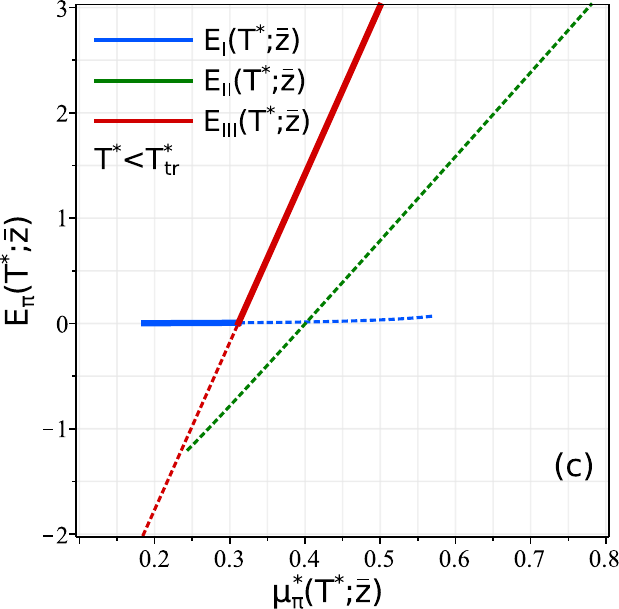}
	\caption{Parametric plots of $E_{I}(T^*;\bar z)$ as a function of the chemical potential $\mu^*_{I} (T^*;\bar z)$ --- blue, $E_{II}(T^*;\bar z)$  versus $\mu^*_{II} (T^*;\bar z)$ --- green, and $E_{III}(T^*;\bar z)$ versus $\mu^*_{III} (T^*;\bar z)$ --- red, with $\bar z$ being a parameter. In Figure~(a), the temperature $T^*=0.15$ is in the range $ T^*_{tr}<T^*<T^*_c $; Figure~(b) shows the special case of the triple-point temperature $T^*=T^*_{tr}=0.135897$; in Figure~(c), the temperature $T^*= 0.125$ is lower than at triple point ($ T^*<T^*_{tr}$). In each figure, bold curve sections display regions of stable phases corresponding to $E_{\pi}(T^*;\bar z_{\rm max})$ and their thin dashed continuations refer to metastable states. The blue, green, and red curves correspond to $\alpha_\pi$ from \eqref{1d19}. Parameters are taken as $f=1.5$ and $v^*=5.0$.	
	}\label{fig3}
\end{figure}

Solutions of systems of equations \eqref{1d30} and \eqref{1d31} are illustrated in Figure~\ref{fig3}a). Here, the parametric plots of functions $E_\pi(T^*;\bar z)$ are shown versus $\mu^*_\pi(T^*;\bar z)$.
The solution of \eqref{1d30} is given by the intersection point of the curves corresponding to $E_{I}$ (blue) and $E_{II}$ (green), which precisely correspond to end-points of their bold sections associated with stable phases (their thin dash continuations refer to metastable states).
The next intersection, of $E_{II}$ (green) and $E_{III}$ (red) corresponds to the solution of \eqref{1d31}. Thus, at subcritical temperatures $T^*<T^*_c$, the system undergoes a sequence of first-order phase transitions between the stable phases: first from Phase~I to Phase~II, and then from Phase~II to Phase~III, where these pairs of phases coexist in equilibrium.

At sufficiently low temperatures (see Figure~\ref{fig3}(c)), a direct phase transition between Phase~I and Phase~III is observed. In this case, the intersection of the $E_{I}$ and $E_{III}$ plots occurs at a lower value of chemical potential than that of $E_{I}$ and $E_{II}$. This behavior indicates the presence of a first-order phase transition between Phase~I and Phase~III associated with the point where the bold curve sections come together. At such low temperatures, the transition between Phase~I and Phase~II becomes thermodynamically inaccessible.
The coordinates of the coexistence points for Phase~I -- Phase~III transition are determined from the system of equations
\begin{equation}\label{1d32}
	\left\{
	\begin{array}{l}
		E_{I}(T^*; z_{I}^{out})=E_{III}(T^*; z_{III}^{in}), \\[1mm]
		\mu_I(T^*; z_{I}^{out})=\mu_{III}(T^*; z_{III}^{in}).
	\end{array}
	\right.
\end{equation}

The analysis of intersection points in Figures~\ref{fig3}(a) and (c) demonstrates that, as the temperature approaches $T^*_{tr}$, the coexistence points related to the phase pairs (I -- II) and (II -- III) converge. At a specific temperature $T^*=T^*_{tr}$, the three curves $E_I(T^*;\bar z)$, $E_{II}(T^*;\bar z)$, and $E_{III}(T^*;\bar z)$ intersect at a single point as displayed in Figure~\ref{fig3}(b). Within this scenario, the extended stability region of the Phase~II in Figure~\ref{fig3}(a)  shrinks to a single point indicated by a full green ring in Figure~\ref{fig3}(b).

At this intersection point, the following set of relations for the functions $E_\pi$ is satisfied:
\begin{equation}\label{1d33}
	\left\{
	\begin{array}{l}
		E_{I}(T^*; z_I^{out})=E_{II}(T^*; z_{II})=E_{III}(T^*; z_{III}^{in}), \\[2mm]
		\mu_I(T^*; z_I^{out})=\mu_{II}(T^*; z_{II})=\mu_{III}(T^*; z_{III}^{in}).
	\end{array}
	\right.
\end{equation}
Under these conditions, the system possesses a \emph{triple point} at $T^*=T^*_{tr}$, where all three stable phases I, II, and III simultaneously coexist in thermodynamic equilibrium.

\section{\label{sec:tp} Triple point on a phase diagram}

\begin{figure}[t!]
	\centering
	\includegraphics[width=0.6\textwidth]{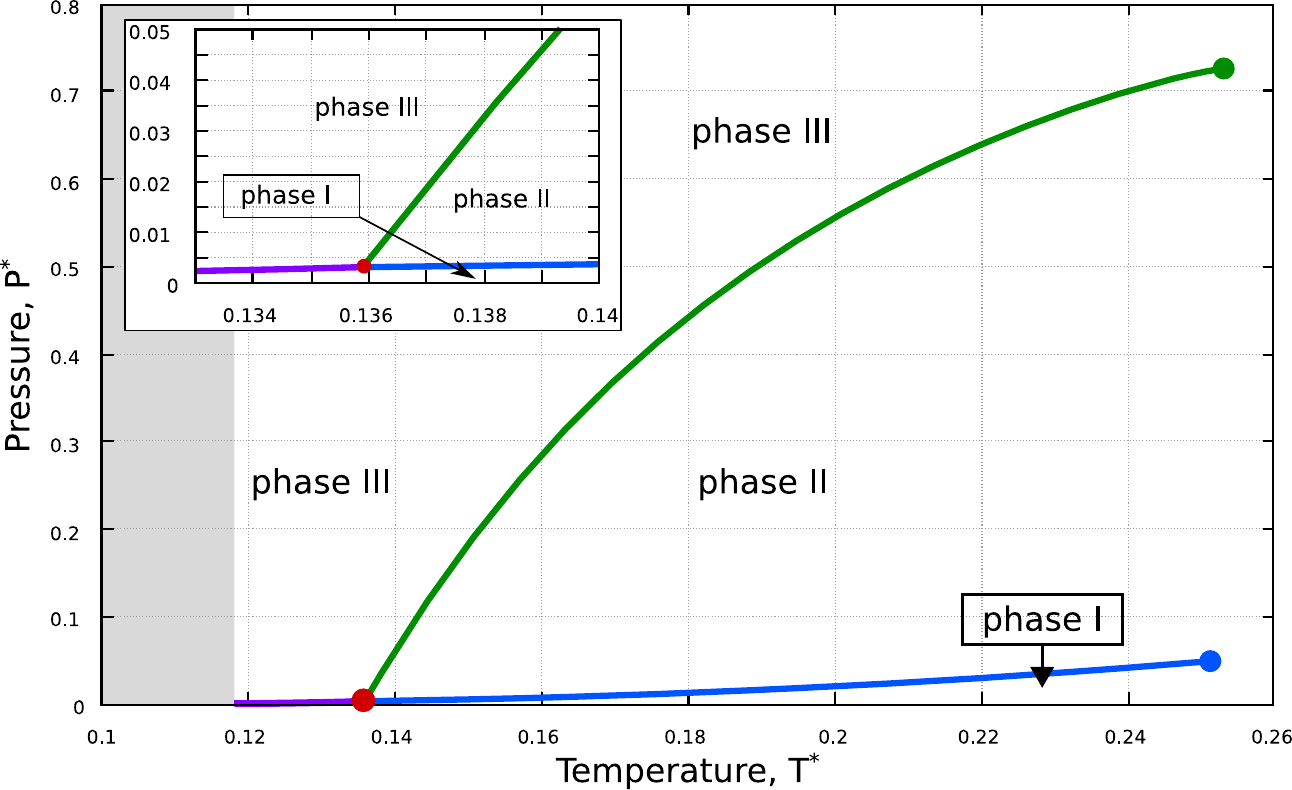}
	\caption{Phase diagram of the CFM with temperature-dependent attraction interaction. The plot spans the temperature interval from $T^*=0.117$ up to the supercritical region (white background). The blue, green and purple curves are the first-order phase transition lines between Phases I, II, and III.
The Phase~I -- II coexistence line is shown in blue, the II --III coexistence line is green, and the I -- III line is purple.
In the high-temperature region, the blue and green curves terminate at their critical points indicated by full circles. The red circle marks the triple point where Phases I, II, and III coexist in thermodynamic equilibrium at a single temperature and pressure. The inset shows the magnified region around the triple point.
As before, $f=1.5$ and $v^*=5.0$.
	}\label{fig4}
\end{figure}

Following the main idea of incorporating an increasing and phase-to-phase dependent attraction part of the interaction potential into the CFM, we
introduced a piece-wise effective chemical potential $\mu^*(T^*;\bar z_{\rm max})$ in \eqref{1d26}, and, using \eqref{1d16} obtained an analogous piece-wise equation of state. This means that for three pases I, II, and III, the pressure $P^* (T^*;\bar z_{\rm max})$ is given by
\begin{equation}\label{1d34}
	P^* (T^*;\bar z_{\rm max})=\left\{
	\begin{array}{l}
		T^* E^*_I (T^*;\bar z_{\rm max}),  \; \text{if} \; \bar z_{\rm max}\in [z_{I}^{in},z_{I}^{out}] \subseteq [0,z_{01}),  \\[1mm]
		T^* E^*_{II} (T^*;\bar z_{\rm max}),  \; \text{if} \; \bar z_{\rm max}\in [z_{II}^{in},z_{II}^{out}] \subseteq  [z_{01}, z_{02}),\\[1mm]
		T^* E^*_{III} (T^*;\bar z_{\rm max}),  \; \text{if} \;  \bar z_{\rm max} \in [z_{III}^{in},z_{III}^{out}] \subseteq  [z_{02}, z_{03}).
	\end{array}
	\right.
\end{equation}

The functions $E^*_{\pi} (T^*;\bar z_{\rm max})$ are influenced by temperature dependence of attraction through the function $\varphi_{\pi}(T^*)$ (see Sec.~ \ref{sec:temp-dep1}) explicitly appearing in \eqref{1d29}.
In particular, a stronger effective attraction (i.e., larger $\varphi_{\pi}(T^*)$) results in a significant negative contribution to the pressure, shifting the coexistence lines downward in the pressure-temperature plane.

Our results are illustrated in Figure~\ref{fig4}, which shows the phase diagram of the modified CFM with temperature-dependent attraction. The diagram spans the temperature interval from the supercritical region $T^*>T^*_c$ down to lowest temperature value satisfying the stability condition $J_2>J_{\pi}(T^*)$ (see Sec.~ \ref{sec:model}). Below this temperature threshold (the gray region in Figure~\ref{fig4}), the attractive interaction becomes dominant over the repulsion, causing the integral in \eqref{1d4} to diverge. The pressure range in Figure~\ref{fig4} covers the region where first two consecutive first-order phase transitions occur in the CFM with constant attraction --- see Figure~\ref{fig1}.

The phase diagram in Figure~\ref{fig4} reflects the interplay between distinct effective temperature-dependent attraction strengths in the three phases involved.
At relatively high temperatures ($T^*_{tr}<T^*<T^*_c$), we observe two first-order transitions related to Phase~I--II and Phase~II--III coexistence lines.
These lines terminate at their critical points. The parameters of these critical points coincide with that of the CFM with constant interaction strengths (see  Figure~\ref{fig1}, Table~\ref{tab1}, which is consistent with our assumption for $J_\pi (T^*)$ in Section~\ref{sec:temp-dep1}.
As soon as the temperature reaches the critical value $T^*_c$, the function $J_\pi (T^*)$ becomes constant for all $T^*\geq T^*_c$.
As the temperature drops below $T^*_{tr}$, the modification of attraction via phase-specific exponents leads to emergence of the Phase~I--III phase boundary.
At the triple point $T^*_{tr}$, all three phases coexist in thermodynamic equilibrium.

Numerical values of the temperature, particle number density, pressure, and chemical potential for the triple point present in the phase diagram given in Figure~\ref{fig4}, are summarized in Table~\ref{tab2}.
At the triple point, all three phases I, II, and III coexist at the same temperature $T^{*}_{tr}$, pressure $P^{*}_{tr}$, and chemical potential $\mu^{*}_{tr}$.
At the same time, the three coexisting phases are characterized by three distinct equilibrium densities $\rho^{*I}_{tr}$, $\rho^{*II}_{tr}$, and $\rho^{*III }_{tr}$.

This is in a full agreement with a qualitative description of the triple point given in the beginning of the Introduction where different densities of coexisting phase have been denoted by $\rho_V$, $\rho_L$, and $\rho_S$ and associated with the vapour, liquid, and solid phases.

\begin{table}[t!]
	\caption{Triple point parameters: the temperature $T^*_{tr}$, three values of the particle number density for distinct phases $\rho^{*I}_{tr}$, $\rho^{*II}_{tr}$, and $\rho^{*III}_{tr}$, pressure $P^*_{tr}$, and chemical potential $\mu^*_{tr}$. For a graphical illustration, see Figure~\ref{fig4}. Additional parameters are $f=1.5$ and $v^*=5.0$.}
	\tabcolsep4.5pt
	\label{tab2}
	\begin{center}
		\begin{tabular}{|c|c|c|c|c|c|}
			\hline
			\multicolumn{6}{|c|}{Triple point} \\
			\hline
			$T^{*}_{tr}$    &   $\rho^{*I}_{tr}$     & $\rho^{*II}_{tr}$     & $\rho^{*III }_{tr}$     & $P^{*}_{tr}$     &   $\mu^{*}_{tr}$    \\
			\hline
			0.135897  &    0.0236806  &    0.976406 &    1.97505  &  0.00297646    &    0.409028  \\
			\hline
		\end{tabular}
	\end{center}
\end{table}

\section{Conclusions}

In this paper, we proposed and analyzed a modified multiple-occupancy CFM with Curie-Weiss-type interactions in which the effective attractive interaction between particles acquires an explicit temperature dependence. While the thermodynamic structure of the original model remains formally intact, the introduction of a temperature-dependent attraction can profoundly alter the topology of the phase diagram. In particular, the modification leads to qualitative changes in the hierarchy of first-order phase transition lines compared to the case of constant interaction strength.

We demonstrated that assigning distinct temperature-dependent attraction strengths to different phases enables the coexistence conditions to be satisfied simultaneously for three phases. This mechanism gives rise to a thermodynamically consistent triple point at a specific temperature $T^*=T^*_{tr}$ and pressure $P^*=P^*_{tr}$, where Phases~I, II, and III coexist in equilibrium with different densities. The triple point is realized when the intersections between the curves representing the functions $E_{I}$, $E_{II}$, and $E_{III}$ (see \eqref{1d29} -- \eqref{1d33}) converge to a single point in the $(E, \mu^*)$ plane as shown in Figure~\ref{fig3}(b).

Our analysis provides a clear physical interpretation of this behavior: as temperature decreases, the relative magnitude of the effective interaction in the higher-density phases increases more rapidly than in the lowest-density phase. This differential temperature dependence reshapes the sequence of first-order phase transition lines, eventually enabling all three phases to share identical pressure and chemical potential. At subcritical temperatures ($T^* < T^*_c$), the resulting phase diagram therefore exhibits a qualitatively new structure compared to the original CFM with constant interactions, where a sequence of infinitely many first-order phase transitions was known to occur but no triple points could be present~\cite{KD22,DKPP26}. In the supercritical region, the  attraction strength remains constant throughout the system, reproducing the behavior of the original CFM.

The proposed modification serves as a minimal and analytically tractable framework in which triple points naturally emerge in a mean-field-type cell fluid model. It illustrates how subtle changes in the temperature scaling of effective interactions can produce nontrivial thermodynamic features, including phase-dependent interactions and intersecting coexistence lines.

\vskip3mm \textit{This work was supported by the National Research Foundation of Ukraine under the project No.~2023.03/0201.}

\vskip3mm \textit{The authors are deeply grateful to all warriors of the Ukrainian Armed Forces, living and fallen, for making this research possible.}





\bibliographystyle{JHEPm}
\addcontentsline{toc}{section}{References}
\providecommand{\href}[2]{#2}
%

\providecommand{\href}[2]{#2}

\end{document}